## Main Manuscript for

# Relief of EGFR/FOS–downregulated miR-103a by loganin alleviates NF-κB–triggered inflammation and gut barrier disruption in colitis

Yan Li[1], Teng Hui[1], Xinhui Zhang[2], Zihan Cao[1], Ping Wang[1], Shirong Chen[1], Ke Zhao[1], Yiran Liu[1], Yue Yuan[1], Dou Niu[1], Xiaobo Yu[1], Gan Wang[1], Changli Wang[3], Yan Lin[4], Fan Zhang[5], Hefang Wu[4], Guodong Feng[6], Yan Liu[7], Jiefang Kang[1], Yaping Yan[1], Hai Zhang[8]*, Xiaochang Xue[1]*, Xun Jiang[4]*

[1]Key Laboratory for Medicinal Resources and Natural Pharmaceutical Chemistry, College of Life Sciences, Shaanxi Normal University; Xi'an, 710119, China.

[2]Department of Digestive, Xianyang Hospital of Yan'an University; Xianyang, 712000, China.

[3]Department of Pathology, Xijing 986 Hospital Department, Air Force Medical University; Xi'an, China.

[4]Department of Pediatrics, Tangdu Hospital, Air Force Medical University; Xi'an, 710000, China.

[5]Department of Neurology, Xianyang Hospital of Yan'an University; Xianyang, 712000, China.

[6]Department of Neurology, Zhongshan Hospital Fudan University; Shanghai, China.

[7]State Key Laboratory of Pathogen and Biosecurity, Changchun Veterinary Research Institute, Chinese Academy of Agricultural Sciences; Changchun, China.

[8]National Translational Science Center for Molecular Medicine and Department of Cell Biology, Air Force Medical University; Xi'an, 710032, China.

*Correspondence to Hai Zhang: (Email: hzhang@fmmu.edu.cn); Xiaochang Xue: (Email: xuexch@snnu.edu.cn); and Xun Jiang: (Email: jiangx@fmmu.edu.cn).

**Keywords:** Ulcerative colitis, miR-103a, loganin, intestinal barrier, EGFR.

**This PDF file includes:** Main Text Figures 1 to 7

## Abstract

Due to the ever-rising global incidence rate of inflammatory bowel disease (IBD) and the lack of effective clinical treatment drugs, elucidating the detailed pathogenesis, seeking novel targets, and developing promising drugs are the top priority for IBD treatment. Here, we demonstrate that the levels of microRNA (miR)-103a were significantly downregulated in the inflamed mucosa of ulcerative colitis (UC) patients, along with elevated inflammatory cytokines (IL-1β/TNF-α) and reduced tight junction protein (Occludin/ZO-1) levels, as compared with healthy control objects. Consistently, miR-103a deficient intestinal epithelial cells Caco-2 showed serious inflammatory responses and increased permeability, and DSS induced more severe colitis in miR-103a$^{-/-}$ mice than wild-type ones. Mechanistic studies unraveled that c-FOS suppressed miR-103a transcription via binding to its promoter, then miR-103a–targeted NF-κB activation contributes to inflammatory responses and barrier disruption by targeting TAB2 and TAK1. Notably, the traditional Chinese medicine *Cornus officinalis* (CO) and its core active ingredient loganin potently mitigated inflammation and barrier disruption in UC by specifically blocking the EGFR/RAS/ERK/c-FOS signaling axis, these effects mainly attributed to modulated miR-103a levels as the therapeutic activities of them were almost completely shielded in miR-103a KO mice. Taken together, this work reveals that loganin relieves EGFR/c-FOS axis-suppressed epithelial miR-103a expression, thereby inhibiting NF-κB pathway activation, suppressing inflammatory responses, and preserving tight junction integrity in UC. Thus, our data enrich mechanistic insights and promising targets for UC treatment.

## Significance Statement

We find that EGFR/RAS/ERK/c-FOS signaling pathway significantly downregulated miR-103a expression in the inflamed mucosa of Ulcerative Colitis (UC) patients and DSS-induced colitis, which then promotes inflammatory response and barrier disruption via activating NF-κB pathway. By means of restoring miR-103a level, loganin, a core ingredient from *Cornus officinalis*, potently attenuates colitis via directly blocking EGFR/RAS/ERK/c-FOS pathway. Thus, targeting EGFR–miR-103a–NF-κB signaling axis serves as an attractive strategy and loganin promises to be a lead compound in UC therapy.

## Introduction

As a systemic autoimmune disorder originating from the intestine, inflammatory bowel disease (IBD) is clinically categorized into two subtypes: ulcerative colitis (UC) and Crohn's disease (CD) based on their distinct clinical manifestations (1). To date, the precise etiology and pathogenesis of IBD are still largely unknown, with the prevailing view attributing it to disrupted immune homeostasis due to the interplays of genetic predisposition, immune imbalance, microenvironment, barrier function, and gut microbiota etc (2-5).

As a classical receptor tyrosine kinase, epidermal growth factor receptor (EGFR) tunes colon epithelial homeostasis from multiple aspects like proliferation, survival, wound repair, barrier maintenance, and ion transport. However, the function of EGFR in IBD has long been paradoxical owing to variable ligands (like EGF, amphiregulin, C26 ceramide, etc.), different cell populations and models (6-8). McCole DF reported that EGFR activation ameliorates diarrhea associated with chronic colitis by improving epithelial sodium absorption (9). Consistently, Zhao JJ and colleagues found that miR-7–targeted EGFR induction alleviates colitis via regulating epithelial cell immunomodulation and regeneration (10). However, Belle NM et al. uncovered that dextran sulfate sodium (DSS) induced more severe colitis in LINGO2KO mice due to EGFR overactivation, while EGFR inhibitor Gefitinib significantly reduced colitis symptoms in LINGO2KO mice (11). We previously found that the classical downstream pathway of EGFR, phosphatidylinositol 3-kinase (PI3K)/protein kinase B1 (AKT1) triggers UC progression via enhancing gut permeability and NF-κB–mediated inflammatory response (12). Therefore, the role and mechanism of EGFR in IBD still deserve further exploration.

Recent studies have shown that dysregulated microRNA (miRNA) closely associated with the onset and progression of inflammation and immune-related diseases (13,14). Specifically, miR-126 is upregulated in the colonic mucosa of UC, where it exacerbates disease progression by suppressing IκBα (15). miR-200b primarily targets ZEB1 to enhance E-cadherin expression, while also promoting the proliferation of intestinal epithelial cells (IECs) to improve barrier function and alleviate colitis (16). miR-103a has been previously reported to play key roles in various cancers (17-20). We uncovered recently that miR-103a was downregulated in the synovial fibroblasts of rheumatoid arthritis (RA) patients by lncRNA H19 sponge adsorption or TNF-α/IL-1β–dependent NF-κB activation, and it then participates in synovial inflammation and bone destruction in collagen-induced arthritis via targeting TAK1, IL-15 and DKK1 (21, 22). Additionally, we found that radiation-suppressed miR-103a promotes glioblastoma (GBM) radiation resistance via inducing cancer stem cell renewal (20). Interestingly, when we generated a whole-body knockout (KO) of miR-103a, the mice showed widespread organ inflammatory responses under steady state, including spontaneous mild colitis, which strongly indicate miR-103a might be a core molecule for the maintenance of immune homeostasis in colitis. However, the relevance of miR-103a and IBD, how miR-103a is regulated and its specific role in IBD remain completely unraveled.

Current IBD therapeutics still face grand challenges. Although substantial progress has been made with immune-targeted biologic drugs, individual response failures remain unsolved (23), and the long-term efficacy and safety are still uncertain (24). Traditional Chinese medicine (TCM), known for its stable therapeutic effects and minimal adverse reactions, has gradually gained attention for IBD treatment (25). *Cornus officinalis* Sieb. et Zucc (CO), a medicinal-edible plant in TCM, is characterized by its sour and astringent taste and slightly warm nature. Emerging evidence has validated huge therapeutic potential of CO in inflammation and oxidative stress reduction, diabetes prevention, liver and neuroprotection (26-28). Its active component loganin (LOG) potently regulates immunocyte activation, reduces inflammatory chemokines, and accelerates Sirt1-mediated p65 deacetylation, possessing good therapeutic effects in DSS-induced colitis (29). However, how CO and its active components alleviate IBD still need elucidation.

Here, we found that miR-103a is a key regulatory node in IBD, as both UC patients and DSS colitis displayed significantly downregulated miR-103a, and DSS induced more severe colitis in miR-103a KO mice compared to wild-type (WT) littermates. Consistently, ethanol extracts of CO apparently attenuated colitis in WT mice, but its effect was potently weakened in miR-103a KO mice. Mechanistically, LOG in CO alleviated colitis via suppressing EGFR/RAS/ERK/c-FOS–miR-103a–NF-κB axis-enhanced gut permeability and inflammatory response. This not only expands the clinical transformation of TCM but also provides new perspectives and targets for the precise intervention of UC.

## Results

### miR-103a is downregulated in the inflamed mucosa of UC patients and acute colitis mice

miR-103a has been previously reported as an anti-inflammatory factor in various diseases like RA (22), and OA (30). To unravel the physiological and pathological significance of miR-103a, we generated miR-103a KO mice with the CRISPR-Cas9 technology (22). The mice exhibited mild spontaneous colitis, suggesting that miR-103a might play a key role in IBD. We hence detected miR-103a levels in the inflamed mucosa of UC patients and DSS-induced colitic mice. Colonoscopy reveals that the intestinal wall of UC patients is rough with granular textures and diffused hyperemia, along with continuous superficial ulcerations (Fig. 1A). Histological analysis demonstrates integrity disruption of the mucosa, disorganized glandular arrangement, significant reduced goblet cells, crypt atrophy accompanied by abscess, and diffuse inflammatory infiltration in the mucosa and submucosa (Fig. 1B). Immunofluorescence (IF) staining shows remarkable downregulation of Occludin, ZO-1, and Villin, indicating damages to the intestinal microvillus structure, disruption of tight junction structures, and severe impairment of the barrier function (Fig. 1C-E). Concurrently, elevated inflammatory cytokines were detected (Fig. 1F). Both RT-qPCR and FISH assays confirmed that miR-103a was significantly downregulated in the inflamed mucosa of UC patients compared to those in the healthy control objects (Fig. 1G-I). Consistently, the DSS-induced colitis mice model was established (Fig. 1J-N) and RT-qPCR results revealed decreased mRNA levels of Occludin and ZO-1 in colitic mice compared to control littermates, along with a significant increase in inflammatory levels *in vivo* (Fig. 1O-P). Notably, miR-103a expression was also markedly downregulated in the colonic tissues of DSS-treated mice (Fig. 1Q). These data indicate that miR-103a might be a key molecule to maintain the hemostasis of immune responses and barrier integrity in intestinal mucosa and thus is involved in colitis development.

### Loss of miR-103a exacerbates intestinal epithelial inflammation and barrier disruption

To validate whether miR-103a affects IBD progression, we built a macrophage CM-induced IECs inflammation model and evaluated the molecule expression profile. As shown in Fig. 2A, CM significantly upregulated IL-1β and TNF-α, along with decreased Occludin, ZO-1, and miR-103a, highly aligns with those observed in UC patients and colitis mice, suggesting that the cell model effectively simulates the pathological phenotypes of UC and is suitable for mechanistic studies. Then, we interfered intrinsic miR-103a level in Caco-2 cells with miR-103a mimics and inhibitor transfection. As expected, RT-qPCR and ELISA detected increased inflammatory cytokines and decreased tight junction proteins in inhibitor-transfected cells. Meanwhile, all these indices were potently reversed in the presence of miR-103a mimics (Fig. 2B, C). IF analysis found that knockdown of miR-103a greatly reduced Occludin/ZO-1 production, but overexpression of miR-103a showed a clear tendency to increase their levels (Fig. 2D, E). These data suggest that miR-103a plays a critical role in the inflammatory response and tight junctions of IECs.

We further investigated the *in vivo* role of miR-103a in DSS-induced colitis. As shown in Fig. 2F, after drinking 2% DSS for 7 days, miR-103a KO mice exhibited a significantly greater weight loss compared to WT mice. Concurrently, KO mice displayed a higher DAI (Fig. 2G), shorter colons (Fig. 2H, I), and heavier spleens (Fig. 2J) during the disease period than the WT mice. Disruption of the mucosal barrier is fundamental to the pathogenesis of colitis (31). IF assessment of the colonic barrier integrity revealed that Occludin, ZO-1 and Villin were more severely downregulated in the mucosa of DSS-treated KO mice compared to WT littermates (Fig. 2K, L). Similarly, AB-PAS staining exhibited heavier intestinal structural damage, fewer goblet cells, and a thinner mucus layer in KO mice than WT mice on DSS insults (Fig. 2M, N). Notably, the mRNA levels of the inflammatory cytokines IL-1β/TNF-α were increased, whereas the tight junction genes were decreased in colon tissues of KO mice compared to those in WT mice, these trends were further enlarged in the presence of DSS (Fig. 2O). Therefore, these data collectively demonstrate that miR-103a deficiency in IECs promotes inflammation and exacerbates intestinal mucosal mechanical and biochemical barrier damage in mice with acute colitis.

**c-FOS activates NF-κB pathway by directly suppressing miR-103a in IECs**

To determine the mechanisms of miR-103a downregulation in UC, Caco-2 cell inflammation model was treated with inhibitors to IBD- and inflammation-related ERK, mTOR, PI3K and Wnt/β-catenin pathways. We found that inhibition of MEK/ERK pathway by PD98059 not only led to the significant decrease of inflammatory factors IL-1β/TNF-α, and the increase of tight junction proteins Occludin/ZO-1, but also completely restored miR-103a reduction (Fig. 3A). In contrast, MEK/ERK pathway agonist TBHQ greatly suppressed miR-103a, which was further strengthened under inflammation conditions, highly compatible with more severe inflammatory response and tight junction disruption (Fig. 3B). ELISA assays confirmed the modulatory effect of the ERK pathway on IL-1β/TNF-α secretion (Fig. 3C, D). Notably, compared to inflammation-induced cells, TBHQ induced a stronger reduction of miR-103a. These data indicate that miR-103a is ERK-dependently downregulated in IECs.

To unravel how ERK regulates miR-103a, we predicted latent transcription factors that may bind miR-103a promoter with transmiR, humanTFDB, and JASPAR databases and dozens of genes were gained (Fig. 3E). Among them, c-FOS, a pivotal gene downstream of ERK, caught our attention (Fig. 3F). We thus conducted luciferase reporter assays by co-transfecting HEK293T cells with pGL3-WT (containing miR-103a promoter sequences) or pGL3-MUT (containing mutant c-FOS binding site), and pCDNA3.1-c-FOS plasmids. As expected, c-FOS potently reduced the luciferase activity in the WT group, but no apparent effect was observed in the MUT group, suggesting that c-FOS inhibits miR-103a by binding its promoter, and ERK downregulates miR-103a via enhancing c-FOS activation (Fig. 3G). Notably, elevated c-FOS was detected in the inflamed mucosa of UC patients (Fig. 3H, and see Supplemental Fig. S1), further consistent with the aforementioned miR-103a downregulation in UC.

We wonder how miR-103a regulates inflammation and mucosal integrity. Signaling pathway interference assays identified that NF-κB mediated miR-103a effects in IECs as NF-κB pathway agonist PMA simulates, but the pathway inhibitor Bay11-7082 reverses the effect of miR-103a on IL-1β/TNF-α/Occludin/ZO-1 expression (Fig. 3I). More importantly, Bay11-7082 but not PD98059 potently reversed miR-103a inhibitor transfection-regulated IL-1β/TNF-α/Occludin/ZO-1 expression (Fig. 3J, K). IF staining of the NF-κB subunit p65 revealed that miR-103a inhibitor alone induced massive p65 nuclear translocation, indicating overactivated NF-κB pathway (Fig. 3L). To confirm miR-103a–targeted NF-κB activation, bioinformatics analysis, luciferase reporter assays (Fig. 3M, see Supplemental Fig. S2A, B), and immunohistochemistry staining (Fig. 3N-Q) identified that miR-103a directly targets TAB2 and TAK1 to trigger NF-κB activation. Thus, these findings suggest that inflammation and barrier dysfunction in UC are driven by miR-103a downregulation, which mediates synergies between ERK and NF-κB pathways.

**CO effectively alleviates colitis by relieving IECs miR-103a suppression**

As miR-103a plays a critical role in inflammation and barrier dysfunction in UC, we screened miR-103a modulators from our TCM library. We found that ethanol extracts of CO strongly elevated miR-103a levels in Caco-2 cells (Fig. 4A). Cytotoxicity assay showed that Caco-2 cells kept intact proliferation under 50 or 100 μg/mL CO stimulation (Fig. 4B). Additionally, CO potently reduced IL-1β/TNF-α, and reversed Occludin/ZO-1 levels, but also greatly restored inflammation-suppressed miR-103a (Fig. 4C). IF staining more intuitively illustrated the effect of CO on tight junction recovery. Compared to the inflammatory model, Occludin/ ZO-1 levels significantly increased after CO treatment (Fig. 4D, E). Not surprisingly, CO effect was potently impaired in miR-103a inhibitor-transfected IECs, suggesting that CO exerts anti-inflammation and barrier repair functions via tuning miR-103a (Fig. 4F-H).

To validate the effect of CO *in vivo*, murine colitis was induced with DSS and 300 mg/kg of CO ethanol extract was administered via gavage simultaneously. As shown in Fig. 4I-M, CO significantly attenuated colitis in WT mice as indicated by improved weight loss, DAI, colon length, and spleen weight. But completely weakened effects in KO mice suggested that miR-103a upregulation mediated CO activity. When gut permeability was measured by oral administration of FITC-Dextran, CO treatment greatly improved intestinal permeability in WT mice ($p < 0.01$), but miR-103a deficiency thoroughly shielded this effect as both CO-untreated and -treated KO mice displayed more severe fluorescence dispersion in the intestines and higher fluorescence leakage in the serum (Fig. 4N, O). Consistently, H&E, AB-PAS, and IF staining of tight junction proteins discovered that DSS intake-caused abnormal mucosal structure, extensive inflammatory infiltration, and goblet cells missing were only mitigated by CO in the WT but not KO mice, which expressed even lower levels of Occludin, ZO-1, and Villin (Fig. 4P-S). Most importantly, RT-qPCR analysis revealed that miR-103a was restored in WT model mice under CO treatment, with reduced inflammatory cytokines and retrieved tight junction genes expression in the colon. But the effect of CO was totally attenuated in the KO mice (Fig. 4T, U). Thus, these findings indicated that CO primarily exerts its therapeutic effects in colitis by upregulating IECs miR-103a.

**CO regulates IECs miR-103a via RAS/ERK signaling pathway**

To unravel how CO regulates miR-103a, the signaling pathways through which CO attenuates UC were predicted by network pharmacology analysis and 207 targets were gained (Fig. 5A). Subsequently, GO and KEGG enrichment analyses were done on these targets with the David database. Results showed that CO components primarily influence protein phosphorylation/ autophosphorylation, and peptidyl-tyrosine phosphorylation in biological processes. As to cellular component regulation, they mainly focus on the plasma membrane, cytoplasm, and receptor complexes. At the molecule level, the functions of them are related to protein kinase activities, and ATP binding (Fig. 5B). The RAS pathway, which is located upstream of ERK/c-FOS pathway, is enriched in the KEGG pathway analysis (Fig. 5C), underlines the probability that CO regulates miR-103a via the RAS/ERK/c-FOS pathway. Hence, we measured the key proteins RAS and ERK in CO-treated Caco-2 cell model and colons of colitis mice. Western blot data revealed that although CO did not significantly alter RAS, phosphorylated ERK (p-ERK) and c-FOS (p-c-FOS) levels in Caco-2 cells under steady state, it did strongly reduce these proteins upregulated by inflammatory induction (Fig. 5D, E). Similarly, RAS and p-ERK proteins were extensively activated in the colons of colitis mice, whereas CO treatment could reverse and downregulate them (Fig. 5F, G). These data indicate that CO might upregulate miR-103a in IECs via the RAS/ERK/c-FOS axis.

**Multi-component in CO synergistically targets EGFR/RAS/ERK pathway**

To identify which components of CO bore its effects on UC, further network pharmacology analysis was done. Based on the STRING database, we constructed a protein-protein interaction (PPI) network and used Cytoscape to screen the top five compounds and the top ten core targets based on degree value, betweenness centrality, and closeness centrality (Fig. 6A-C). Building on previous data, we hypothesize that CO may directly target EGFR, hence causing downstream RAS/ERK pathway abnormal transduction. Molecular docking was used to assess the binding affinity of CO-derived 7-O-methylmorroniside, LOG, hydroxygenkwanin (HGK), ethyl oleate, cornudentanone, ethyl linolenate and tetrahydroalstonine to EGFR and PyMOL software was used for visualization (Fig. 6D, and see Supplemental Fig. S3). Notably, a binding energy of ≤ -5 kcal/mol generally indicates a good binding interaction between the docked molecules. As shown in Fig. 6C, the calculated binding energies of these compounds were all below -5 kcal/mol, and they formed various covalent bonds with EGFR. Therefore, CO might exert its therapeutic effects via multiple bioactive small molecules targeting EGFR.

To validate the predictions, LOG and HGK were selected for *in vitro* verification. LOG, an iridoid glycoside, is a well-studied anti-inflammatory monomer abundant in CO (32). HGK, a flavonoid, exhibits anti-inflammatory and anti-tumor effects (33, 34). Bio-layer interferometry (BLI) assay revealed that LOG and HGK bind to EGFR with a sub-micromolar affinity constant ($K_D$) of 0.74 nM and 0.053 nM, respectively (Fig. 6E, L). Additionally, the cellular thermal shift assay (CETSA) confirmed that LOG directly binds and inhibits EGFR degradation (Fig. 6F). Through screening of their superior efficacy in modulating inflammatory cytokine levels, tight junctions, and miR-103a expression, we determined the optimal concentrations for subsequent experiments to be 100 µg/mL for LOG and 10 µM for HGK (Fig. 6G, M). IF staining and ELISA confirmed that both LOG and HGK effectively suppressed inflammatory cytokine secretion and restored tight junction proteins in the inflammation model (Fig. 6H, I, N, O, P). Consistently, LOG and HGK significantly suppressed CM-induced abnormal activation of EGFR/RAS/ERK pathway in Caco-2 cells as indicated by falling levels of EGFR, p-EGFR, RAS, and p-ERK (Fig. 6K, Q, R). These data suggested that CO might attenuate UC via directly targeting EGFR/RAS/ERK axis by the components LOG and HGK.

**LOG alleviates DSS-induced colitis via the EGFR/RAS/ERK pathway**

To validate LOG's therapeutic efficacy and mechanism in DSS colitis *in vivo*, we administered LOG at doses of 50,100, and 200 mg/kg to WT model mice via gavage, with 5-ASA as the positive control drug. We found that the LOG-M group (100 mg/kg) exhibited the most optimal therapeutic effects, even surpassing those of the positive control, as demonstrated by the most significant recovery in body weight, reduction in DAI, as well as increased colon length and decreased spleen coefficient (Fig. 7A-E). H&E staining confirmed that LOG treatment significantly ameliorated DSS-induced colonic structural damage (Fig. 7F). Notably, RT-qPCR assays showed that LOG substantially upregulated miR-103a in the inflamed mucosa *in vivo* (Fig. 7G). Western blot also indicated that LOG markedly downregulated EGFR level and inhibited the activation of the EGFR/RAS/ERK pathway (Fig. 7H, I). Thus, these data fully validated that LOG exerted itself as a core active component to mediate the therapeutic effects of CO on UC via regulating the EGFR/RAS/ERK and NF-κB signaling pathways, between which miR-103a plays a pivotal bridging role (Fig. 7J).

## Discussion

We demonstrated that miR-103a stands in the center position between the EGFR/RAS/ERK and NF-κB pathways, and loss of it in the inflamed mucosa triggered severe inflammatory responses and barrier dysfunction in UC. Colitis mice with miR-103a deficiency exhibited exacerbated disease severity, whereas CO and its active component LOG alleviated DSS colitis via regulating miR-103a expression by specifically targeting the EGFR/RAS/ERK pathway. Therefore, our data provide a novel and promising strategy for IBD treatment.

Although miR-103a has been verified to play pivotal roles in various diseases like cerebral ischemia reperfusion injury (35), RA and cancer (19, 22), its expression and functions in colitis are almost completely unknown. We uncovered that miR-103a was significantly downregulated in the intestinal mucosa of DSS-induced colitis, and its deficiency directly exacerbated barrier disruption and inflammatory responses. Zhao et al. reported that Berberine attenuates murine colitis via upregulating miR-103a and hence suppressing Wnt/β-catenin pathway activation, but how miR-103a was regulated in colitis is not mentioned (36). Our data uncovered that miR-103a is downregulated in UC patients and DSS colitis in a c-FOS-dependent manner and is involved in EGFR/RAS/ERK/c-FOS and NF-κB pathway synergistic regulation, by which it extensively modulates inflammatory response and barrier damage in colitis.

Due to its dominant regulation of cell proliferation and metastasis, researches about EGFR mainly focus on the field of cancer (37). Accumulating evidence demonstrated that EGFR tunes intestinal epithelial homeostasis through various patterns, but the concrete function of EGFR in colon has long been paradoxical owing to variable ligands, different distribution and complex gut microenvironment (6-8). For example, many studies have shown that EGFR can promote epithelial cell regeneration to repair intestinal barrier damage, but long-term EGFR activation often leads to an increased risk of cancer (10, 38-40). For colitis, both beneficial and harmful effects of EGFR-associated pathway were reported (9-11). In our study, EGFR was overexpressed and overactivated in the IECs of colitis, and disease remission was observed when LOG and HGK targeted and inhibited EGFR activation. While this finding appears contradictory to some literature, it may relate to oxidative stress—a key driver of gut hyperpermeability in IBD (41). $H_2O_2$-induced oxidative stress disrupts tight junctions in Caco-2 cells via activating the EGFR/ERK pathway (42). Consistently, LOG inhibits $H_2O_2$-induced neuronal apoptosis by blocking ERK pathway (43), and HGK suppresses the proliferation and migration of non-small-cell lung cancer by enhancing EGFR degradation (44). This further supports the notion that EGFR may be a critical regulatory site for CO. Therefore, we speculate that fluctuations in EGFR levels may be attributed to different stages of IBD. Although we focused on EGFR as a core target, other highly correlated targets predicted by network pharmacology (such as AKT1, SRC, etc.) still require further exploration to determine their involvement in the effects of CO. For CO-origin EGFR inhibitor selection, although LOG displays lower affinity to EGFR than HGK, it more weakly affects EGFR activation under the steady state, indicating a safer status, and lower affinity allows LOG to dissociate with EGFR more flexibly and hence block more EGFR ligands. Absolutely, we can't exclude the superior effect of HGK over LOG on colitis.

Immune-inflammatory regulation has long been a central focus in IBD pathogenesis. The NF-κB family, a core regulator of immune-inflammatory responses, is strongly implicated in IBD via dysregulated signaling (36, 45, 46). Zhang and our lab discovered that miR-103a can regulate NF-κB pathway by targeting TAK1, thereby inhibiting inflammation in RA (22, 47), which aligns with our findings. Regarding the downstream mechanisms of miR-103a, we verified that it regulates NF-κB pathway by targeting TAB2 and TAK1, and thus exacerbating inflammation and tight junction disruption. Remarkably, miR-103a KO mice exhibited lower levels of Occludin/ZO-1 in colonic tissues compared to WT controls under steady state, implying that miR-103a maintains intestinal barrier integrity by regulating tight junction protein expression. This discovery explains the molecular basis of miR-103a downregulation in UC progression, and also provides a theoretical foundation for miRNA-targeted interventions.

Although CO has been listed in the Chinese medicines with the same origin as medicine and food by Chinese National Health Commission in 2018 after clinically validating for over 2000 years for tonifying the liver and kidneys, its clinical application prospects in the intestines are far from elucidated. We have fully demonstrated the strong beneficial effects of loganin on both acute and chronic DSS colitis (another article under review), but several critical steps are still warranted before it can be translated into clinical application, either individual or joint. First, extensive preclinical development is essential, including pharmacokinetic and pharmacodynamic studies and toxicity assessment. Second, investigational new drug (IND)-enabling studies should be conducted to define a safe starting dose and identify biomarkers for target engagement and efficacy monitoring. Finally, well-designed clinical trials in patients with UC would be necessary to evaluate the safety and efficacy in UC treatment (a single-center double-blind clinical trial is under preparation in Tangdu hospital).

In summary, CO relieves c-FOS–restrained miR-103a via braking EGFR expression and activation, the elevated miR-103a then suppresses TAB2 and TAK1 expression to prevent NF-κB activation, and ultimately inhibits the inflammatory response and tight junctions' disruption. This study uncovered the complex synergies between EGFR and NF-κB pathways, and validated the rationality of treating UC via targeting the EGFR/RAS/ERK/c-FOS–miR-103a–NF-κB axis. We also demonstrate the clinical application prospects of TCM in the treatment of UC, offering critical insights for reducing disease relapse and advancing the clinical translation of natural therapeutics.

# Figures

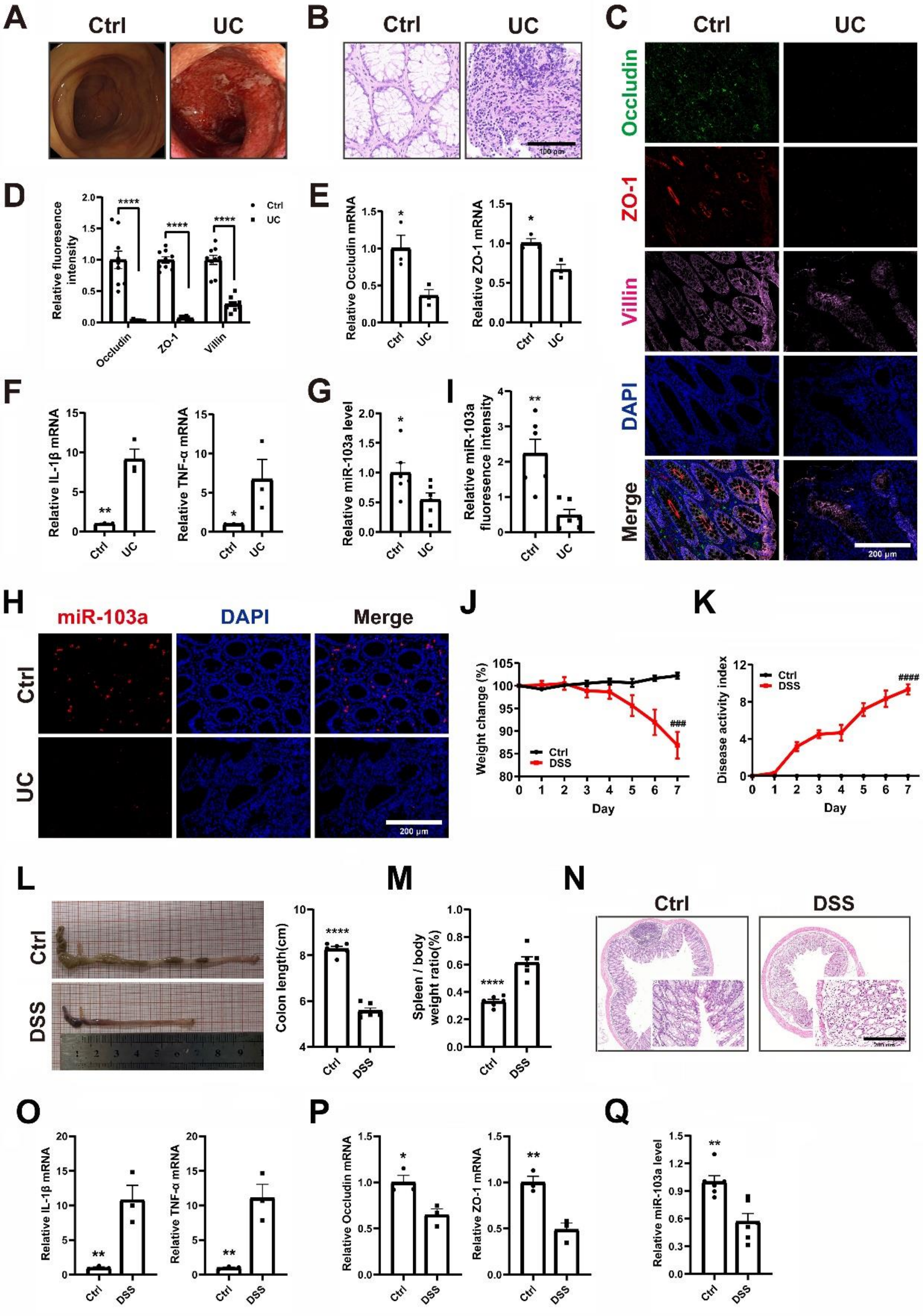

**Fig. 1. miR-103a is significantly downregulated in UC patients and DSS-induced murine colitis.** (**A**) Colonoscopic images of UC patients and healthy controls. (**B**) H&E staining of colonic biopsies reveals destruction of crypt structures and massive inflammatory infiltration in UC. Scale bar, 100 μm. (**C-D**) Immunofluorescence staining of Occludin, ZO-1, and Villin expression in colonic biopsy sections of UC and control (C) and quantification analysis (D). Scale bar, 200 μm. (E-G) mRNA levels of proinflammatory cytokines, tight junction proteins and miR-103a in colon biopsies were measured by RT-qPCR ($n = 3$). (**H-I**) FISH staining of colonic biopsies to detect the levels of miR-103a. Scale bar, 200 μm. (**J-K**) Experimental colitis was induced in WT C56BL/6J mice ($n = 6$) via 2% DSS in drinking water for 7 days. Body weight changes are presented relative to day 0 (set as 100%) (J), and daily DAI was recorded (K). (**L**) Representative colon images and statistical analysis of colon length on day 7 ($n = 6$). (**M**) The ratio of spleen weight to body weight on day 7 are shown. (**N**) Representative H&E-stained paraffin sections of mice colons on day 7. Scale bar, 200 μm. (**O-Q**) mRNA levels of proinflammatory cytokines, tight junction proteins ($n = 3$), and miR-103a ($n = 6$) in mice colon tissues on day 7, analyzed by RT-qPCR. Data are presented as mean ± SEM. Statistical significance was determined by unpaired t-test or one-way ANOVA. $^*p < 0.05$, $^{**}p < 0.01$, $^{****}p < 0.0001$ vs UC or DSS group, and $^{\#}$ vs Ctrl group. Ctrl, control; DSS, dextran sulfate sodium.

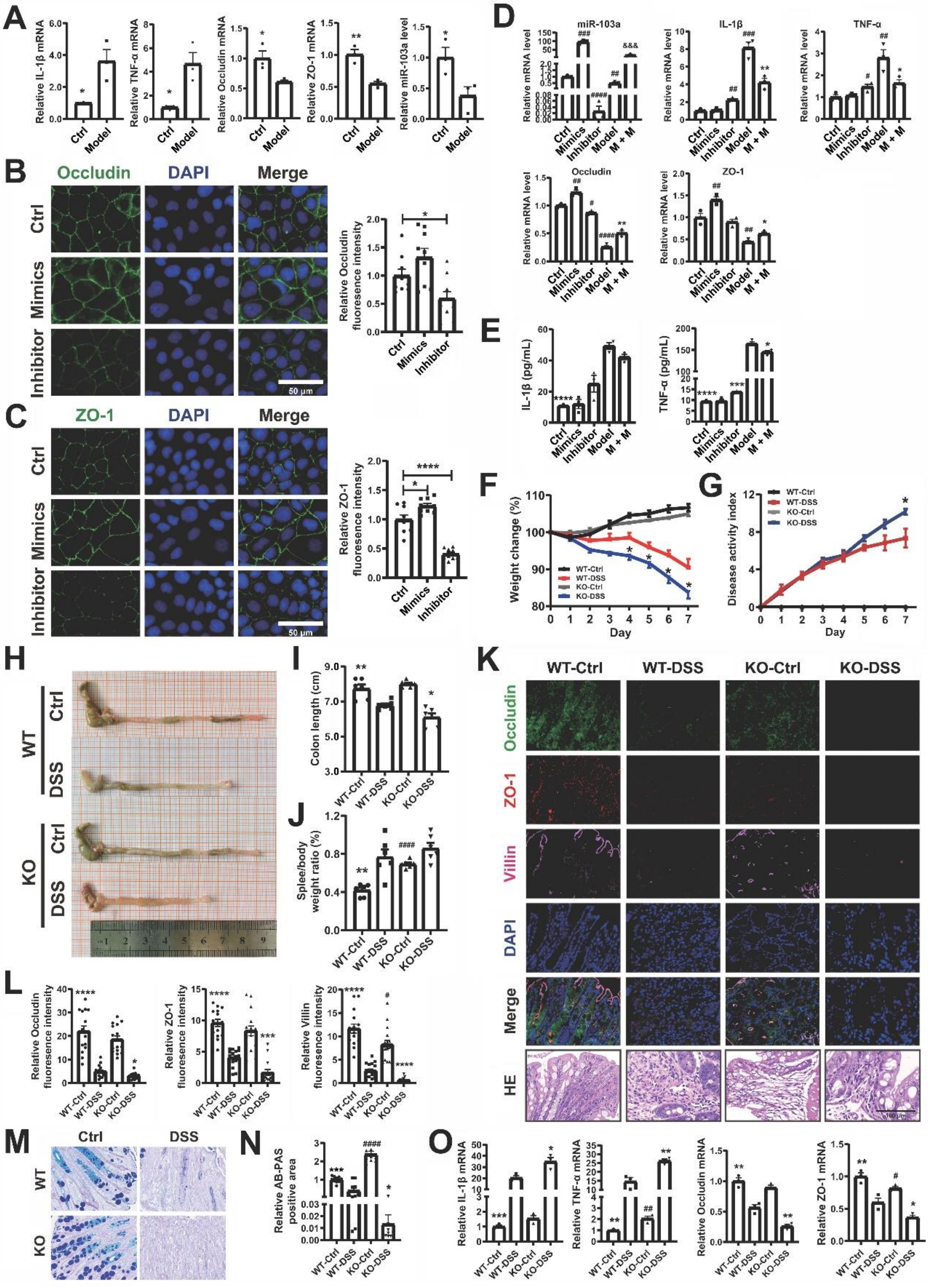

**Fig. 2. Loss of miR-103a exacerbates colitis and promotes inflammation and barrier disruption.** (A) THP-1 cells were stimulated with LPS (100 ng/ml) for 24 h, and the supernatant was collected as CM. Caco-

2 cells were treated with 50% CM for 24 h to establish an inflammatory model. The mRNA levels of proinflammatory cytokines, tight junction proteins, and miR-103a were measured by RT-qPCR (*n* = 3). (**B-C**) Caco-2 cells were transfected with miR-103a mimics or inhibitor, and IF was used to detect the levels of tight junction proteins Occludin (B) and ZO-1 (C) (*n* = 3). (**D**) Caco-2 cells were transfected or treated as indicated for 24 h, the levels of miR-103a, proinflammatory cytokines and tight junctions were detected by RT-qPCR (*n* = 3), $^{\&}$ indicates comparison with the mimics group. (**E**) IL-1β and TNF-α levels in cell supernatants were measured by ELISA (*n* = 3). (**F-G**) Colitis was induced in WT and KO mice with 2% DSS, body weight (F) and the DAI (G) changes were recorded daily (*n* = 6). (**H-J**) Representative colon tissue images were taken (H), colon length (I) and the ratio of spleen weight to body weight (J) were statistically analyzed. (**K-L**) Levels of Occludin, ZO-1, and Villin in colon sections were detected by IF (K) and statistically analyzed (L). Bar, 100 μm. (**M-N**) Goblet cell numbers were assessed by AB-PAS staining (M) and analyzed (N). (**O**) Proinflammatory cytokines and tight junction proteins levels in colon tissues were measured by RT-qPCR (*n* = 3). Data are presented as mean ± SEM. Statistical significance was determined by unpaired t-test or one-way ANOVA. $^{*}p < 0.05$, $^{**}p < 0.01$, $^{***}p < 0.001$, $^{****}p < 0.0001$. * vs Model or WT-DSS group, and $^{\#}$ vs Ctrl or WT-Ctrl group. CM, conditioned medium; M + M, model + mimics; IF, immunofluorescence staining; WT, wild type; KO, knockout; DSS, dextran sulfate sodium.

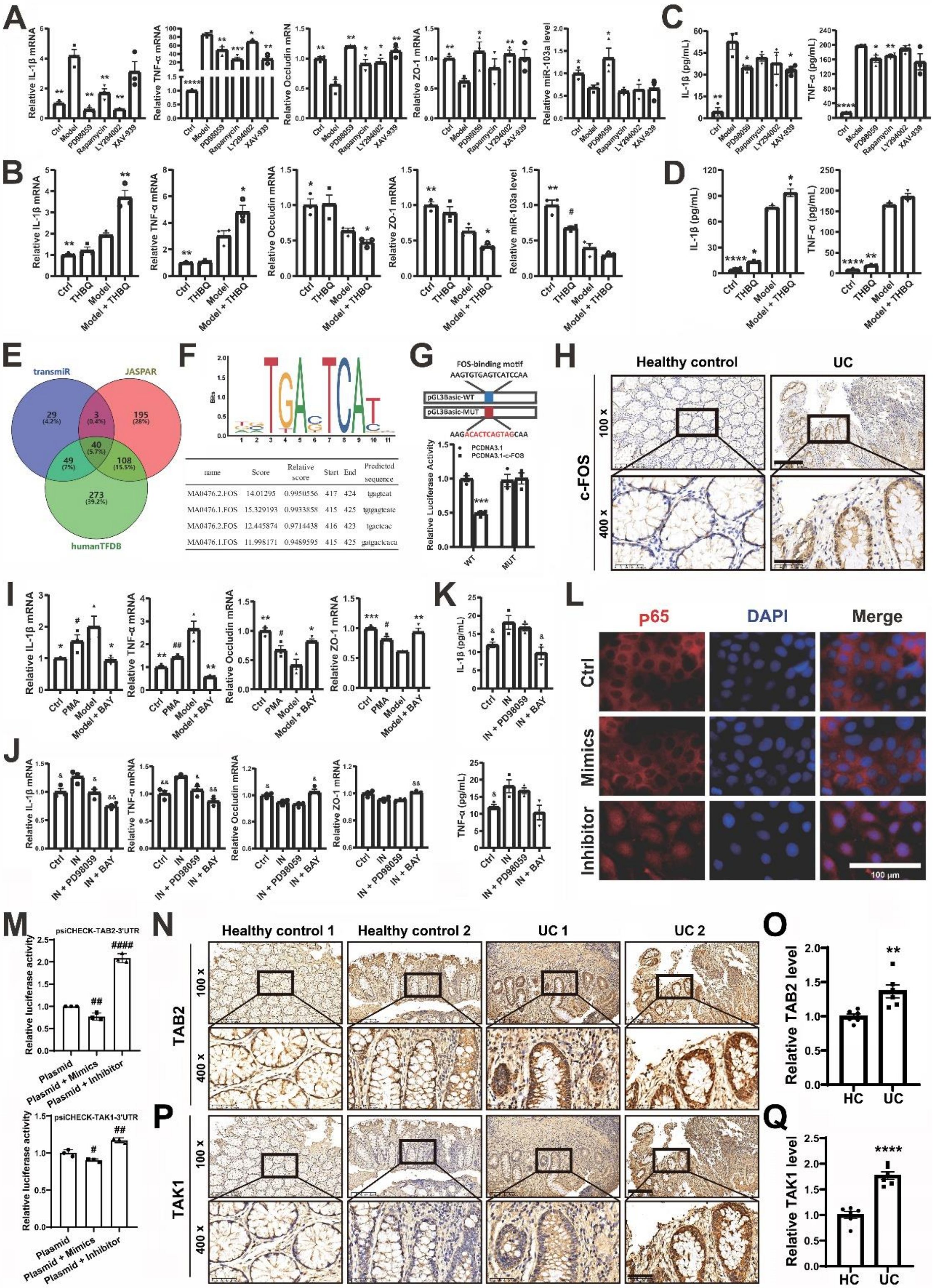

**Fig. 3. miR-103a deficiency exacerbates colitis by promoting inflammation and barrier disruption.** (**A-D**) Caco-2IM was treated with inhibitors (10 μM) targeting ERK, mTOR, PI3K, and Wnt/β-catenin (A, C)

or ERK (B, D) pathways for 24 h. RT-qPCR (A, B) and ELISA (C, D) were used to measure proinflammatory cytokines, tight junction proteins, and miR-103a levels ($n$ = 3). (**E**) Potential miR-103a–regulating TFs were predicted by bioinformatics analysis. (**F**) Potential binding sites of c-FOS in miR-103a promoter were analyzed with JASPAR. (**G**) HEK293T cells were co-transfected with recombinant pGL3Basic plasmids containing WT or MUT c-FOS binding sequences in miR-103a promoter and pcDNA3.1-c-FOS, and luciferase activities were measured 24 h later. (**H**) Expression of c-FOS in colon biopsies was detected by IHC. Scale bars, 200 μm for 100×magnification and 50 μm for 400×magnification. (**I-J**) Caco-2 cells were treated with the NF-κB pathway agonist PMA or Caco-2 IM was treated with inhibitor BAY11-7821 (I), or Caco-2 cells were transfected with miR-103a inhibitor and treated with PD98059 and BAY11-7821 (J), the levels of proinflammatory cytokines and tight junctions were detected by RT-qPCR 24 h later ($n$ = 3). 10 μM for each chemical reagent. (**K**) IL-1β/TNF-α levels were measured by ELISA ($n$ = 3). (**L**) Caco-2 cells were transfected as indicated and p65 nuclear translocation was detected by IF. (**M**) Dual-luciferase reporter assays were used to confirm miR-103a target genes. (**N-Q**) TAB2 and TAK1 levels in colon biopsies of UC and healthy controls were detected by IHC and statistically analyzed. Scale bars, the same as in H. Data are presented as mean ± SEM. Statistical significance was determined by unpaired t-test or one-way ANOVA. $^{*}p < 0.05$, $^{**}p < 0.01$, $^{***}p < 0.001$, $^{****}p < 0.0001$. * vs Model group, $^{\#}$ vs Ctrl group. Ctrl, control; TFs, transcription factors; WT, wild type; MUT, mutant; BAY, BAY11-7821; IN, inhibitor; Caco-2IM, Caco-2 inflammatory model; HC, healthy control; UC, ulcerative colitis.

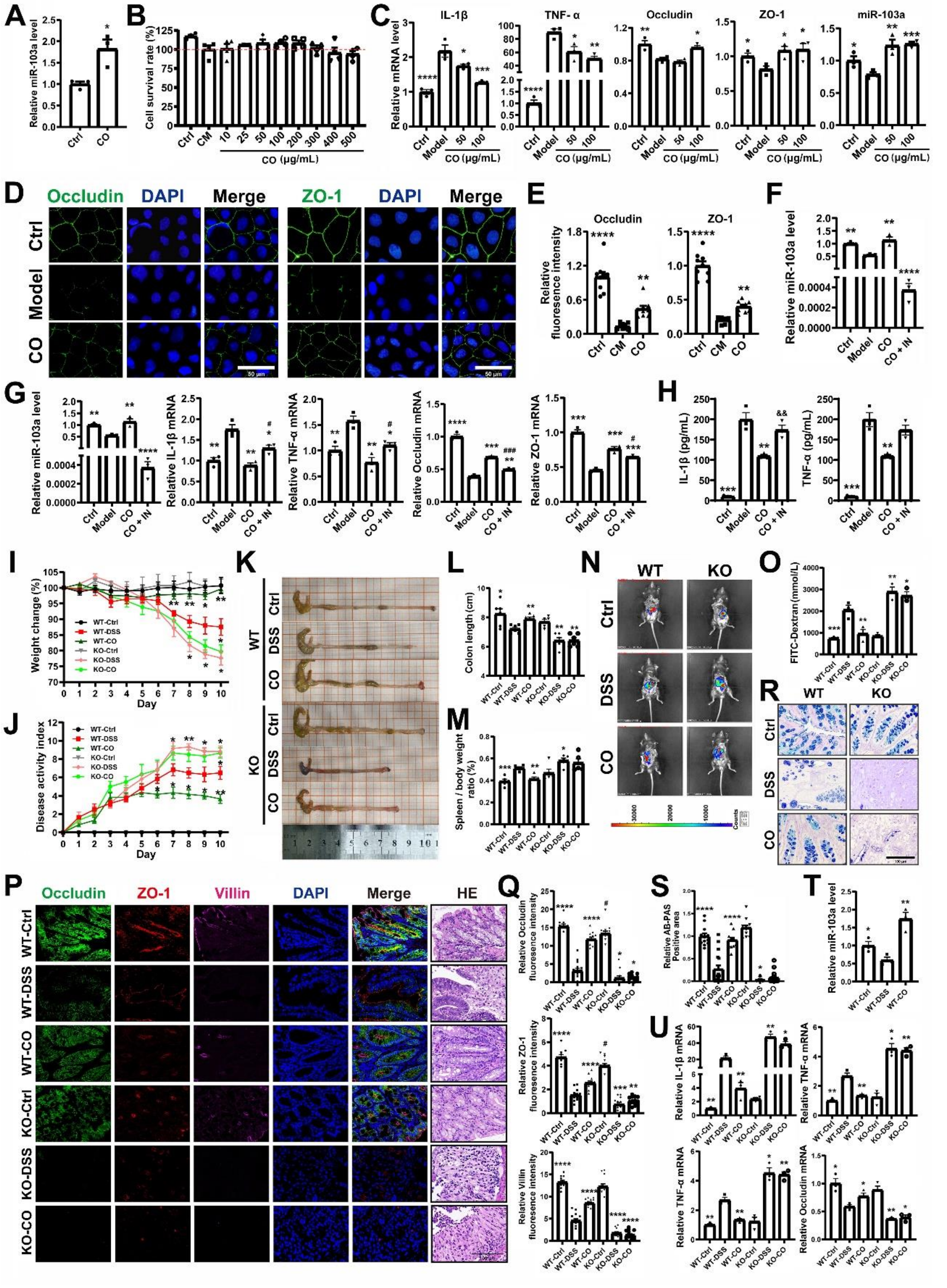

**Fig. 4. CO ethanol extract alleviates DSS colitis by modulating miR-103a.** (**A**) Caco-2 cells were stimulated with 200 μg/mL CO ethanol extract for 24 h, and the mRNA of miR-103a was detected by RT-

qPCR (*n* = 3). (**B**) Caco-2IM was treated with serial-diluted CO ethanol extract for 24 h. Cell proliferation was assessed via CCK8 assay (*n* = 4). (**C**) Caco-2IM were stimulated with 50 and 100 μg/mL CO ethanol extract, and various gene levels were detected by RT-qPCR (*n* = 3). (**D-E**) CO ethanol extract (100 μg/mL) was used to treat Caco-2IM for 24 h and Occludin and ZO-1 expression was detected by IF (D) and analyzed (E) (*n* = 8). Bar, 50 μm. (**F-H**) miR-103a inhibitor was transfected into Caco-2 cells, then Caco-2IM was induced followed by CO ethanol extract treatment for 24 h (*n* = 3). RT-qPCR was used to detect the levels of genes (F, G), and ELISA was used to detect IL-1β and TNF-α levels (H), $^{\&}$ vs the CO group. (**I-M**) Experimental colitis was induced in WT and miR-103a KO mice (*n* = 6) via 2% DSS for 7 days, followed by 3 days of normal diet. Mice in the CO group were administered 300 mg/kg CO ethanol extract prepared with 0.5% CMC-Na daily. Changes in Body weight (I) and DAI (J) were recorded daily, representative colon images (K) and statistical analysis of colon length (L) and the ratio of spleen to body weight (M) were obtained on day 10. (**N-O**) On the 10th day, after fasting for 8 h, the mice were administered 400 mg/kg of FITC-Dextran. 6 h later, *in vivo* fluorescence imaging (N) was performed on the abdominal cavity, and serum was collected to measure fluorescence intensity (O) (*n* = 3). (**P-Q**) Occludin, ZO-1, and Villin expression in mice colon tissues were detected by IF (P)and statistically analyzed (Q). (**R-S**) Goblet cell quantification via AB-PAS staining on day 10. (**T-U**) On the 10th day, various gene levels in the colon tissues were detected by RT-qPCR (*n* = 3). Data are presented as mean ± SEM. Statistical significance was determined by unpaired t-test or one-way ANOVA. **p* < 0.05, ***p* < 0.01, ****p* < 0.001, *****p* < 0.0001. * vs Model or WT-DSS group, $^{\#}$ vs Ctrl group. Caco-2IM, Caco-2 inflammatory model; CM, conditioned medium; Ctrl, control; IN, miR-103a inhibitor; WT, wild type; KO, knockout; DSS, dextran sulfate sodium.

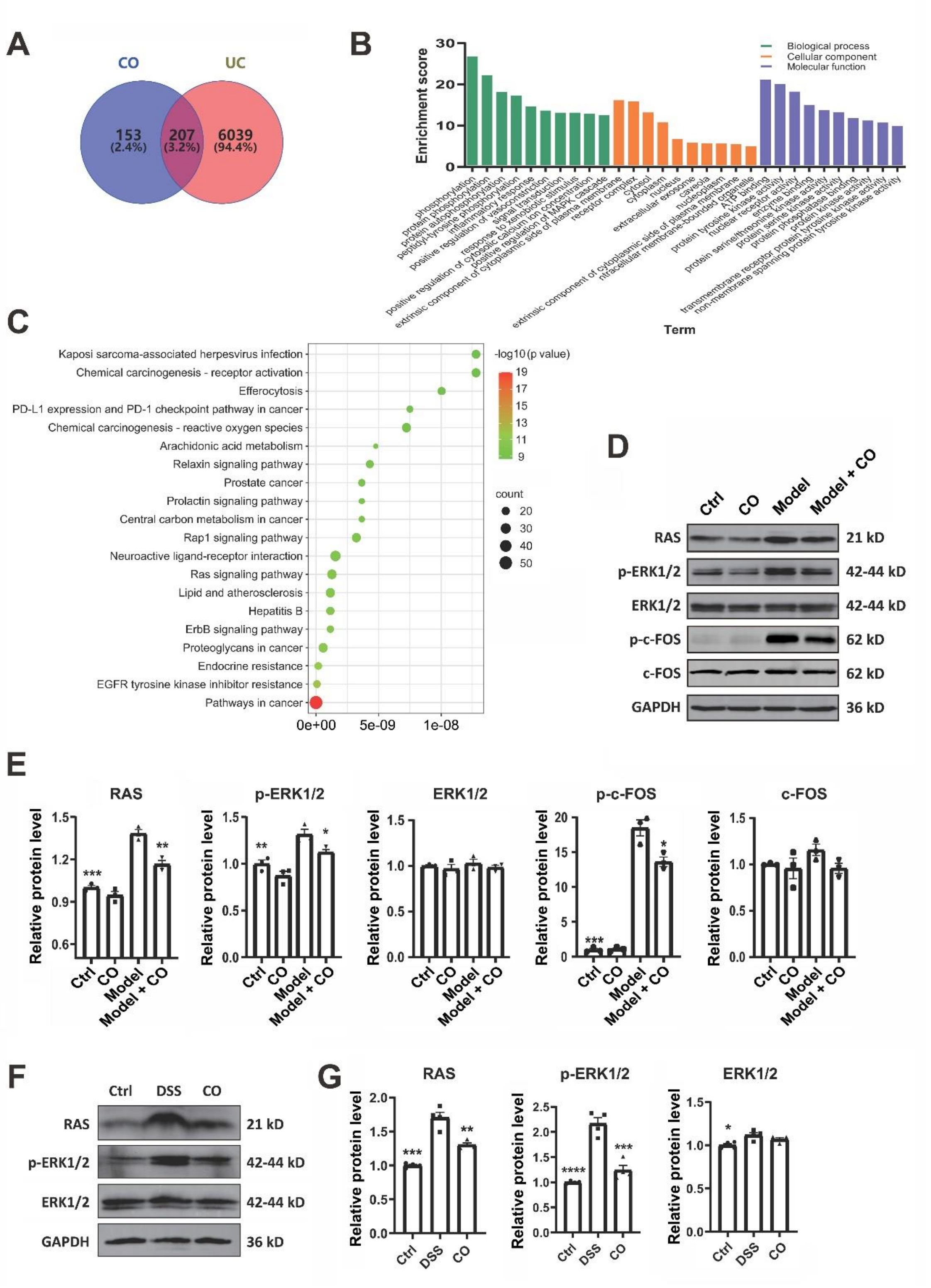

**Fig. 5. Network pharmacology predicts the potential mechanisms of CO in UC treatment.** (**A**) Venn diagram showing overlapping targets between CO and UC. (**B**) GO enrichment analysis. (**C**) KEGG signaling

pathways. (**D-E**) Western blot analysis of RAS, p-ERK, ERK, p-c-FOS and c-FOS protein expression in Caco-2 cells (*n* = 3) treated with 100 μg/mL CO ethanol extract for 24 h. (**F-G**) Experimental colitis was induced in WT mice via 2% DSS for 7 days, followed by 3 days of normal diet. The CO group of mice was administered 300 mg/kg CO ethanol extract prepared with 0.5% CMC-Na daily. Immunoblot analysis of RAS, p-ERK, and ERK protein expression levels (*n* = 3). Data are presented as mean ± SEM. Statistical significance was determined by unpaired t-test or one-way ANOVA. *$p < 0.05$, **$p < 0.01$, ***$p < 0.001$, ****$p < 0.0001$. * vs Model or WT-DSS group. Ctrl, control; DSS, dextran sulfate sodium.

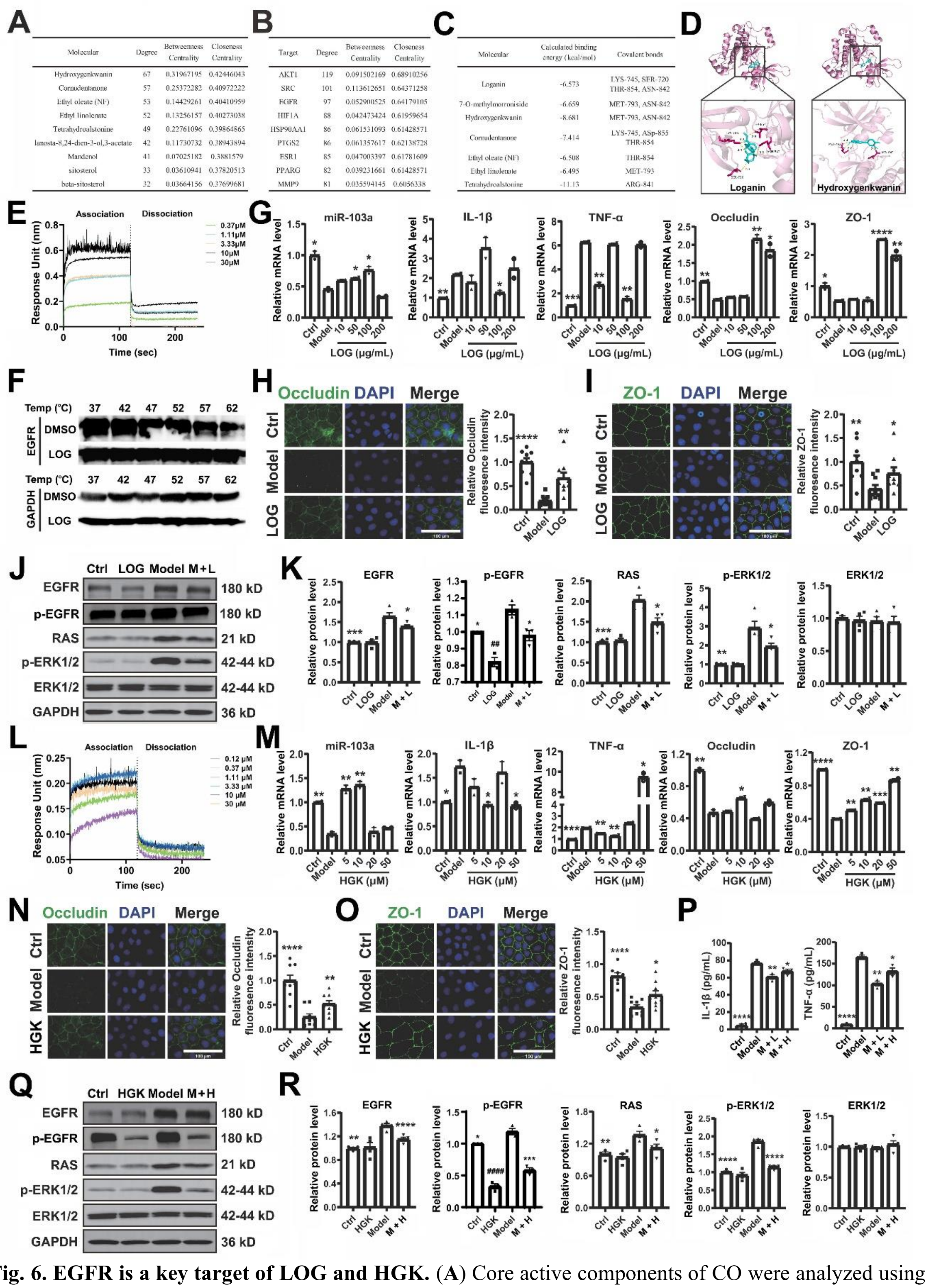

**Fig. 6. EGFR is a key target of LOG and HGK.** (**A**) Core active components of CO were analyzed using Cytoscape. (**B**) Core therapeutic targets of CO were identified via STRING database analysis. (**C**) Prediction

of the docking binding energy between active components and EGFR using AutoDockTools. (**D**) Docking schematic diagram drawn by PyMOL. (**E, F, L**) Direct binding of LOG (E, F) or HGK (L) with EGFR was detected by BLI (E, L) and CETSA (F). (**G, M**) Inflammatory Caco-2 cells ($n$ = 3) treated with serial-diluted LOG (G) or HGK (M) for 24 h, and the levels of various genes were analyzed by RT-qPCR. (**H, I, N, O**) Caco-2IM was stimulated with 100 μg/mL LOG (H, I) or 10 μM HGK (N, O) for 24 h, followed by detection of the expression of Occludin (H, N) and ZO-1 (I, O) by IF and statistically analyzed ($n$ = 3). (**J, K, Q, R**) Western blot analysis of EGFR, RAS, p-ERK, and ERK protein expression in LOG- (J, K) or HGK- (Q, R) treated cells ($n$ = 3). (**P**) Caco-2IM was treated with LOG or HGK for 24 h and IL-1β and TNF-α secretion levels were analyzed by ELISA ($n$ = 3). Data are presented as mean ± SEM. Statistical significance was determined by unpaired t-test or one-way ANOVA. *$p < 0.05$, **$p < 0.01$, ***$p < 0.001$, ****$p < 0.0001$. * vs Model group, $^{\#}$ vs Ctrl group. Caco-2IM, Caco-2 inflammatory model; Ctrl, control; LOG, loganin; HGK, hydroxygenkwanin; BLI, bio-layer interferometry; CETSA, cellular thermal shift assay; M + L, model + LOG; M + H, model + HGK.

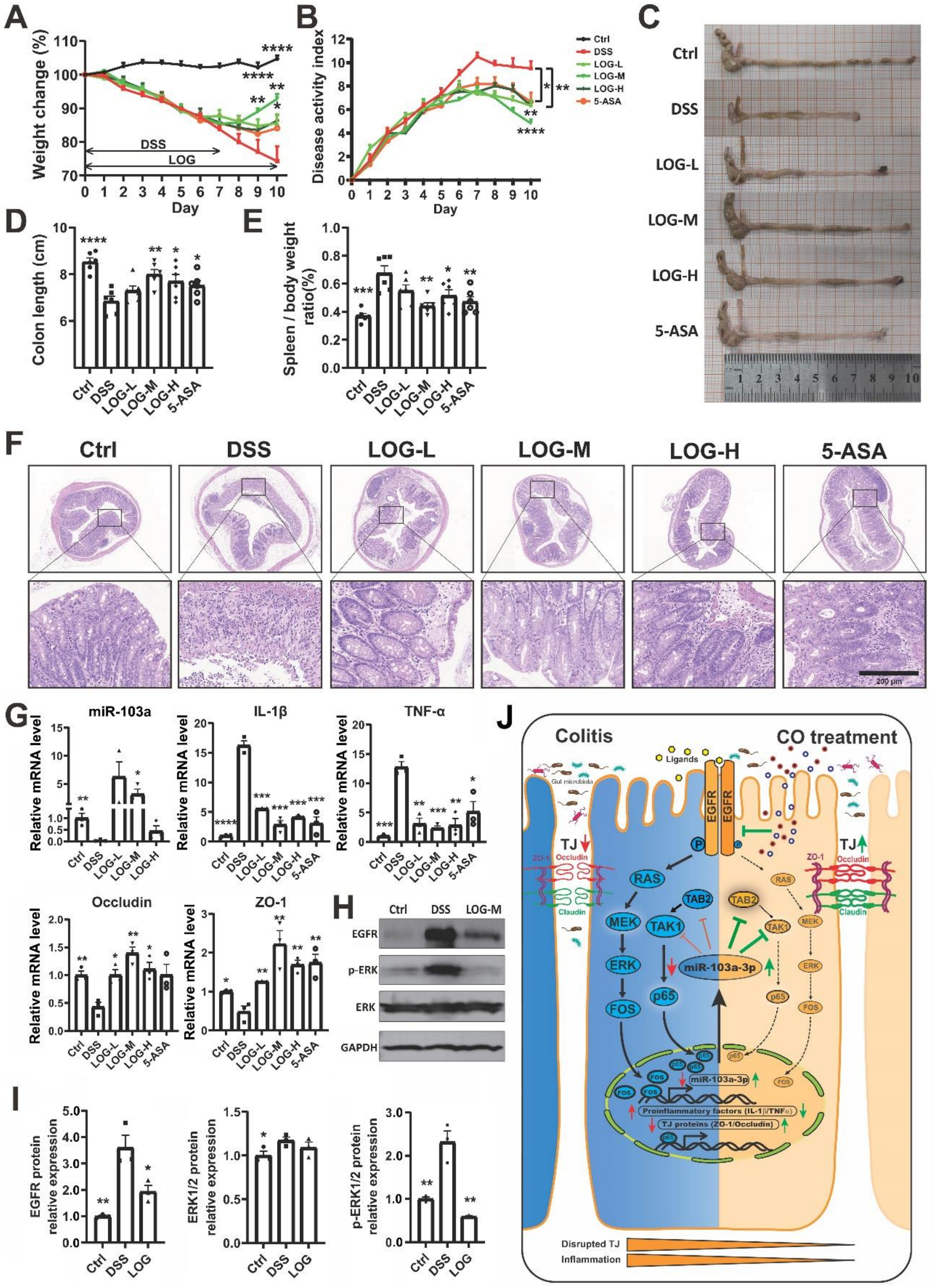

**Fig. 7. LOG alleviates DSS-induced colitis via EGFR signaling.** (**A-B**) Experimental colitis was induced in WT mice (*n* = 6) via 2% DSS for 7 days, followed by 3 days of normal diet. The LOG group of mice was administered 50 mg/kg, 100 mg/kg, and 200 mg/kg LOG, respectively, each prepared in 0.5% CMC-Na, daily, while the 5-ASA group received 150 mg/kg 5-ASA prepared in 0.5% CMC-Na daily. Changes in body weight are presented relative to day 0 (set as 100%) (A), and DAI was recorded daily (B). (**C-E**) On day 10, representative colon images (C), statistical analysis of colon length (D), and the spleen-to-body weight ratio (E) were documented ($n$ = 6). (**F**) Histopathological examination of colon sections via H&E staining. (**G**) On day 10, the mRNA levels of miR-103a, proinflammatory cytokines, and tight junctions in mice colon tissues were detected by RT-qPCR ($n$ = 3). (**H-I**) Western blot analysis of EGFR, p-ERK, and ERK protein expression ($n$ = 3). (**J**) graphical abstract. Data are presented as mean ± SEM. Statistical significance was determined by unpaired t-test or one-way ANOVA. *$p < 0.05$, **$p < 0.01$, ***$p < 0.001$, ****$p < 0.0001$. * vs DSS group. Ctrl, control; DSS, dextran sulfate sodium; DAI, disease activity index.

## Materials and Methods

**Mice.** The WT C57BL/6 mice were from the Experimental Animal Center of Shaanxi Normal University, and miR-103a KO mice were provided by Prof. Hai Zhang from the Air Force Medical University. All mice were routinely fed in specific pathogen-free lab at 25°C with a 12-h light/dark cycle. All animal experiments were approved by the Animal Ethics Committee of Shaanxi Normal University and conducted in accordance with the committee's guidelines.

**Patients.** All UC patients were recruited from Tangdu Hospital of the Air Force Medical University. After clinical confirmation, colonic biopsies were collected from the lesioned areas of patients or normal areas of healthy controls according to requirements and preserved in fixative solution. This study was approved by the Ethics Committee of Tangdu Hospital, and informed consent was obtained.

**Histology.** On day 10, mice were euthanized with 1% sodium pentobarbital. Distal colons were fixed in 4% paraformaldehyde for 1-2 days, followed by dehydration and paraffin embedding. Tissue sections were then stained with hematoxylin and eosin (H&E) or Alcian blue-periodic acid-Schiff (AB-PAS) for histological examination of the colon.

**Cell culture.** Caco-2 human colorectal adenocarcinoma cells, THP-1 human monocytic cells, and HEK293T human embryonic kidney cells were obtained from ATCC. Caco-2 cells were cultured in MEM medium (Procell, PM150410) supplemented with 1% Non-essential Amino Acid (Meiluncell, PWL088) and 1 mM sodium pyruvate solution (Procell, PB180422). THP-1 cells were grown in RPMI 1640 medium (Thermo Fisher, 11875093). HEK293T cells were cultured in DMEM high-glucose medium (Mishushengwu, MI00622). All media were supplemented with 10% fetal bovine serum (EXCELL, FSP500), 100 U/mL penicillin, and 100 μg/mL streptomycin (Mishushengwu, MI00614). Cells were maintained in an incubator at 37°C and 5% $CO_2$.

**Cell transfection.** Transfection of miR-103a mimics (RiboBio, miR10000101-1-5), inhibitor (miR20000101-1-5) and plasmids into cells was performed using Lipofectamine 2000 (Invitrogen, 11668019). Briefly, cells were seeded in 12-well plates at a density of $2 \times 10^5$ cells/well and cultured until reaching 70–80% confluency., The cells were transfected following the manufacturer's protocol and cultured for 24-48 h before further processing.

**Murine acute colitis model induction and treatment.** Both WT and miR-103a KO male C57BL/6 mice (6-8-week-old) were randomly divided into normal control, DSS (36-50 kDa, MP Biomedicals) model, and drug treatment groups (≥6/group). The control mice drank distilled water, all others were given water containing 2% DSS for 7 days, followed by 3 days of distilled water. The treatment groups were orally administered with solutions of CO alcohol extract or LOG, with CMC-Na as the solvent, from the first day of model induction, and treated for 10 consecutive days. The control and model group received equivalent CMC-Na solution.

**Pathological scoring.** As previously mentioned (12), DAI was daily evaluated during modeling through body weight changes, stool consistency, and fecal occult blood.

**Establishment of the intestinal epithelial cell inflammation model.** THP-1 cells were seeded at a concentration of $2 \times 10^6$ cells/mL and treated with 100 ng/mL PMA for 48 h to induce M0 macrophages. Then, the cells were incubated with complete MEM containing 100 ng/mL LPS overnight. After removing cell debris by centrifugation (1200 rpm × 5 min), the supernatant was mixed with an equal volume of MEM medium to make a 50% conditioned medium (CM). When Caco-2 cells reached 70% confluency, the medium was replaced with 50% CM and cultured overnight to establish the IECs inflammation model.

**Reverse transcription quantitative polymerase chain reaction (RT-qPCR).** Total RNA was extracted from tissues or cells by RNAiso Plus (TaKaRa, 9109), and cDNA was synthesized with reverse transcription cDNA kits (Vazyme, R223-01). For miR-103a detection, cDNA was synthesized using a stem-loop RT primer with miRNA-specific reverse transcription kits (Vazyme, MR101). Gene expression levels were analyzed using ChamQ SYBR qPCR Master Mix (Vazyme, Q311-02) on the StepOnePlus real-time PCR instrument. miR-103a expression level was detected using the miRNA dye-based qPCR Mix kit (Vazyme, MQ102). The primers used were listed in Table S1. β-Actin or housekeeping miRNA U6 was used as internal controls, and gene expression levels were calculated with the $2^{-\Delta\Delta Ct}$ method.

**Fluorescence in situ hybridization (FISH).** The expression level of miR-103a in colonic tissue specimen was detected with FISH assay as previously described (22). A specific fluorescence-labeled has-miR-103a probe (5'-TCATAGCC CTGTACAATGCTGCT-3'), was designed and synthesized by Servicebio (Wuhan, China). The experiment was performed according to the protocol provided by Servicebio.

**Western blot.** Proteins were extracted from cells and tissues using RIPA lysis buffer supplemented with PMSF and phosphatase inhibitors. Then, the protein samples were quantified using a BCA protein assay kit (Mishushengwu, MI00606B), and heated with 5 × loading buffer at 95°C for 10 min. After separated by SDS-PAGE, the proteins were transferred to PVDF membrane followed by a block with 5% skimmed milk. Then, the membrane was incubated with specific primary and peroxidase-labeled secondary antibodies at 4°C, and protein bands were visualized using a chemiluminescence imaging system, with GAPDH as the internal control. Finally, protein expression was quantitatively analyzed using Image J software.

**Enzyme-linked immunosorbent assay (ELISA).** Cell culture supernatants were collected and centrifuged at 3500 rpm for 10 min. The clarified supernatants were then subjected to ELISA for quantification of IL-1β and TNF-α according to the manufacturer's protocols. Absorbance was measured using a microplate reader, and cytokine concentrations were determined from standard curves.

***In vivo* imaging and intestinal permeability assay in mice.** After 10 days of model induction, three mice from each group were randomly selected for intestinal permeability assessment. Following an 8-h fasting period, FITC-Dextran (Sigma-Aldrich, FD4) with a molecular weight of 3000-5000 Da was administered via oral gavage at a concentration of 400 mg/kg and blood was collected 6 h later. The mice were then euthanized and their abdominal cavities were exposed. The distribution of FITC-Dextran was visualized using an IVIS SpectrumCT *In Vivo* Imaging System. Concurrently, plasma was obtained and the fluorescence intensity of FITC-Dextran was measured as well.

**Immunofluorescence.** Caco-2 cells were seeded on cell culture slides at a concentration of $2 \times 10^5$ cells/mL, treated as indicated, and fixed by immersion in 4% PFA. Paraffin-embedded tissue sections underwent antigen retrieval in EDTA buffer. Both cells and tissues were incubated in blocking buffer containing 0.2% Triton X-100, 5% FBS, and 3% bovine serum albumin (BSA) at 37°C for 1 h. Then, the samples were incubated with primary antibodies at 4°C overnight, followed by labeling with fluorescence-conjugated secondary antibodies. Finaly, cell nucleus was counterstained with DAPI (Acbam, ab104139) for 5 min, and images were captured using an Olympus fluorescence microscope (CKX3-MVR).

**Luciferase reporter assay.** The predicted binding sequence within the promoter region of miR-103a was inserted into the pGL3-Basic plasmid. To assess the specificity of the interaction between c-Fos and miR-103a, HEK293T cells were seeded in a 12-well plate at a density of $2 \times 10^5$ cells/mL and co-transfected with the pGL3-Basic plasmid containing the miR-103a promoter sequence, and the c-Fos overexpression plasmid. The empty pGL3 plasmid was used as a negative control. Luciferase activity was measured 48 h post-transfection. For miR-103a target gene validation, dual luciferase reporter assay was used. Briefly, both TAB2 and TAK1 3'UTR oligonucleotide fragments were cloned into psiCheck2 plasmid and were used to co-transfect HEK293T cells with miR-103a mimics or inhibitor, and *Renilla* and Firefly luciferase activities were measured 24 h later with Dual-Luciferase® Reporter Assay System (Promega, E1910).

**Network pharmacology.** Active components of CO were screened in the TCMSP database (https://old.tcmsp-e.com/tcmsp.php) based on oral bioavailability (OB) ≥ 30% and drug-likeness (DL) ≥ 0.18. Then, chemical components were added and integrated by reviewing the literature. Convert all active components into SMILES format using the PubChem database (https://pubchem. ncbi.nlm.nih.gov/), and predict potential targets of each component via SwissTargetPrediction (http://swisstargetprediction.ch/). Relevant targets of UC were retrieved and summarized by the DisGeNET (https://disgenet. com/) and GeneCards databases ((https://www.genecards.org/). The obtained targets were imported into the Venny 2.1.0 website (https://bioinfogp.cnb.csic.es/ tools/venny/) to identify the intersection of drug and disease target genes. To further explore the potential mechanisms of drug treatment for UC, the intersection targets were subjected to GO and KEGG enrichment analysis in the DAVID database (https://david.ncifcrf.gov/). Select the top 10 items with the smallest *p*-values for biological processes (BP), cellular components (CC), molecular functions (MF), and the top 20 items for KEGG pathways.

**Molecular docking.** The compounds loganin, 7-O-methylmorroniside, hydroxygenkwanin, cornudentanone, ethyl oleate (NF), ethyl linolenate, and tetrahydroalstonine were retrieved from the TCMSP database, while the protein structure of EGFR_4I24 was downloaded from the PDB database. Molecular docking was performed using AutoDockTools software, and the results were visualized with PyMOL.

**Cell viability assay.** The therapeutic effect of CO was evaluated using the CCK-8 assay. Caco-2 cells were seeded in 96-well plates at a density of $5 \times 10^3$ cells/well and cultured for 24–48 h. Subsequently, the cells were treated with serially-diluted CO alcohol extract (0, 10, 25, 50, 100, 200, 300, 400, 500 μg/mL) for 24

h. Then, 10 μL of CCK-8 reagent was added to the cells and incubated at 37℃ for 4 h. The absorbance at 495 nm was finally measured using a microplate reader.

**Bio-layer interferometry (BLI) binding assay.** The binding affinities of LOG and HGK with EGFR (Zeye Biotechnology Ltd., China) was determined by using BLI on Octet RED (FortéBio, Shanghai, China) as previously described (48). In brief, all the analyses were carried out at 30℃ in 0.2% DMSO/PBS buffer. Loading of Ni-NTA biosensors was done by exposing EGFR (Catalog Number ZY60248HuP) containing 0.1 mg/mL to biosensor tips for 2 h. Then, a 96-well plate was filled with 200 μL of sample or buffer per well and shook thoroughly at 1000 rpm. The loaded biosensors were washed in buffer for 10 min and transferred to the wells containing serially-diluted LOG or HGK as indicated. Both the association and dissociation were observed for 1 min for each sample. Buffer alone was used as reference measurements, and Ni-NTA unloaded biosensors was used as a control. The sample-sensorgrams were normalized through subtracting the double reference curve. Binding affinity ($K_D$) were obtained from a global fit to a 1:1 binding model of the data between EGFR and LOG/HGK with Octet software (FortéBio).

**Cellular thermal shift assay (CETSA).** To validate LOG target engagement in Caco-2 cell lysates, a CETSA assay was performed. Briefly, lysates were obtained from $2 \times 10^6$ Caco-2 cells, diluted with cold PBS, and divided into aliquots. Each aliquot was then mixed with LOG (10 μM) or DMSO in eppendorf tubes and incubated for 0.5 h at room temperature, followed by heating to indicated temperatures with a Veriti thermal cycler (Life Technologies, USA). Finally, aggregated proteins were removed and the supernatants were subjected to standard Western blot analysis with EGFR antibody.

**Statistical analysis.** All experiments were repeated at least three times, with data expressed as mean ± standard error of the mean (SEM). Statistical analyses were conducted using GraphPad software, with comparisons between groups conducted using unpaired t-tests or one-way ANOVA. A $p$-value of $< 0.05$ was considered statistically significant.

**Acknowledgments**

We thank Zhaoqiang Qian, Lifang Zheng and Qiangqiang Wei from Laboratory Animal Center of Shaanxi Normal University for their support and assistance in animal feeding, management and experiment.

**Author Contributions**

G.D.F., Y.L., J.F.K., Y.P.Y., X.C.X., and X.J. are responsible for the concept and design. Y.L., T.H., X.H.Z., Z.H.C., P.W., S.R.C., K.Z., Y.R.L., Y.Y., D.N., X.B.Y., G.W., C.L.W., Y.L., F.Z., G.D.F., Y.L., J.F.K., and Y.P.Y. performed *in vitro* experiments. Y.L., T.H., and X.H.Z. conducted *in vivo* experiments. X.H.Z., X.J., and H.F.W. collected colonic tissue specimen. H.Z. generated the miR-103a KO mice. J.F.K., H.Z., X.C.X., and X.J. applied for the grants. Y.L., Z.H.C., and X.C.X. analyzed the data and wrote the paper.

**Competing Interest Statement**

Authors declare that they have no competing interests.

**Funding**

This work was supported by Key Research and Development Projects of Shaanxi Province (No. 2024SF-ZDCYL-03-07 and 2024SF-ZDCYL-04-16). Open Fund Project of Key Science and Technology Innovation Platform of Central Universities (No. GK202205010 and GK202407001). Joint Funds of the National Natural Science Foundation of China (No. U24A20787). National Natural Science Foundation of China (No. 82270563). Special Fund for Military Laboratory Animals (SYDW_KY (2021)13).

**Ethics approval**

All procedures with human tissue specimens were performed after approval of the Ethics Committee of the Air Force Medical University and written informed consent (202203-138), and animal experiments were approved by the Animal Ethics Committee of Shaanxi Normal University (202512065).

**Patient consent**

Obtained.

## Supplementary Figures

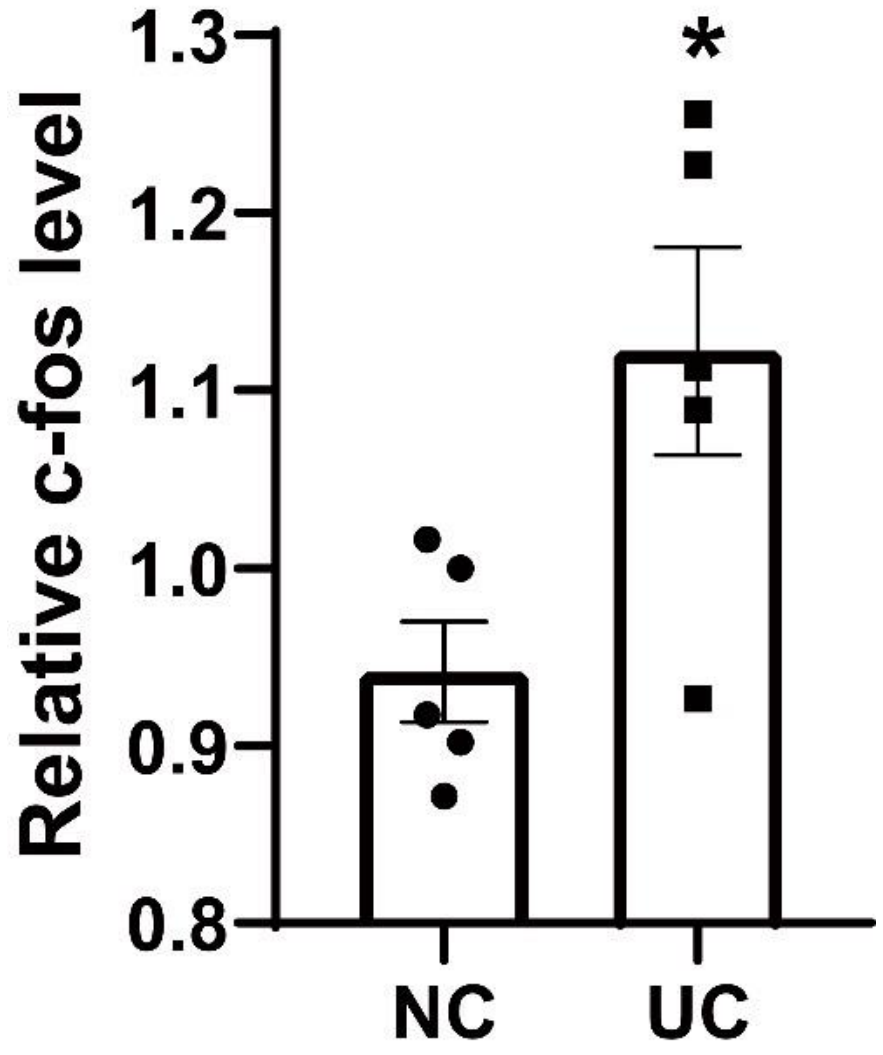

**Fig. S1. Quantification of immunohistochemistry staining of c-FOS in colon tissues of patients with UC.**

Data are analyzed by unpaired Student's t-test (*$p < 0.05$).

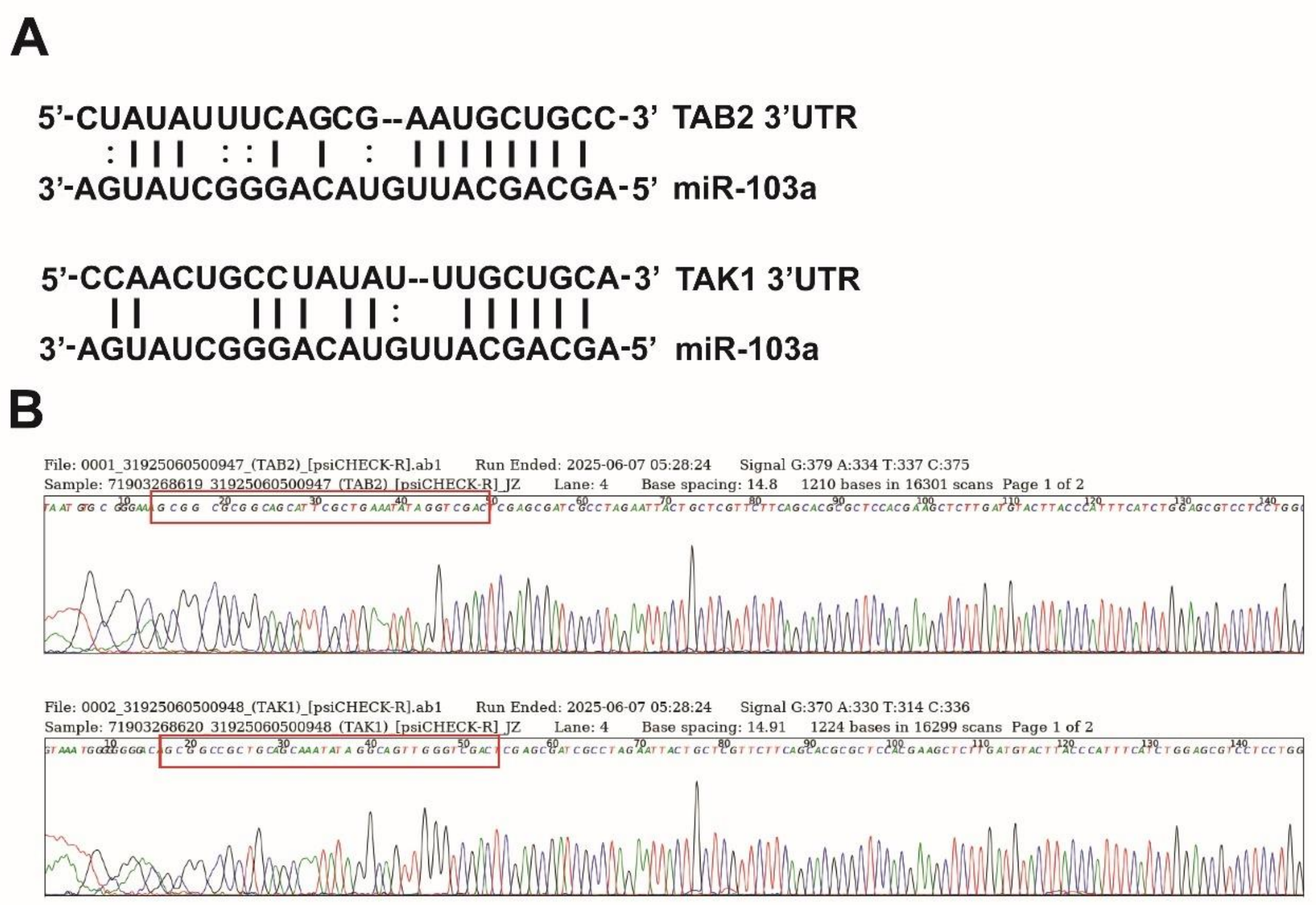

**Fig. S2. Sequence alignment of miR-103a and target genes and dual-luciferase reporter plasmids sequencing.** (**A**) miR-103a target genes were screened by bioinformatics analysis with ENCORI (https://rnasysu.com /encori/index.php) and RNA sequences alignment were performed between mature miR-103a-3p and the seed sequence-matched 3'UTR of candidate target genes TAB2 and TAK1. (**B**) DNA sequencing results of the recombinant dual-luciferase reporter plasmids which contain the 3'UTR of TAB2 and TAK1 between Not I (GCGGCCGC) and Sal I (GTCGAC) endonucleases sites of psiCheck2 plasmid. Red box indicates the correct sequence of 3'UTR of TAB2 and TAK1 cloned into psoCheck2.

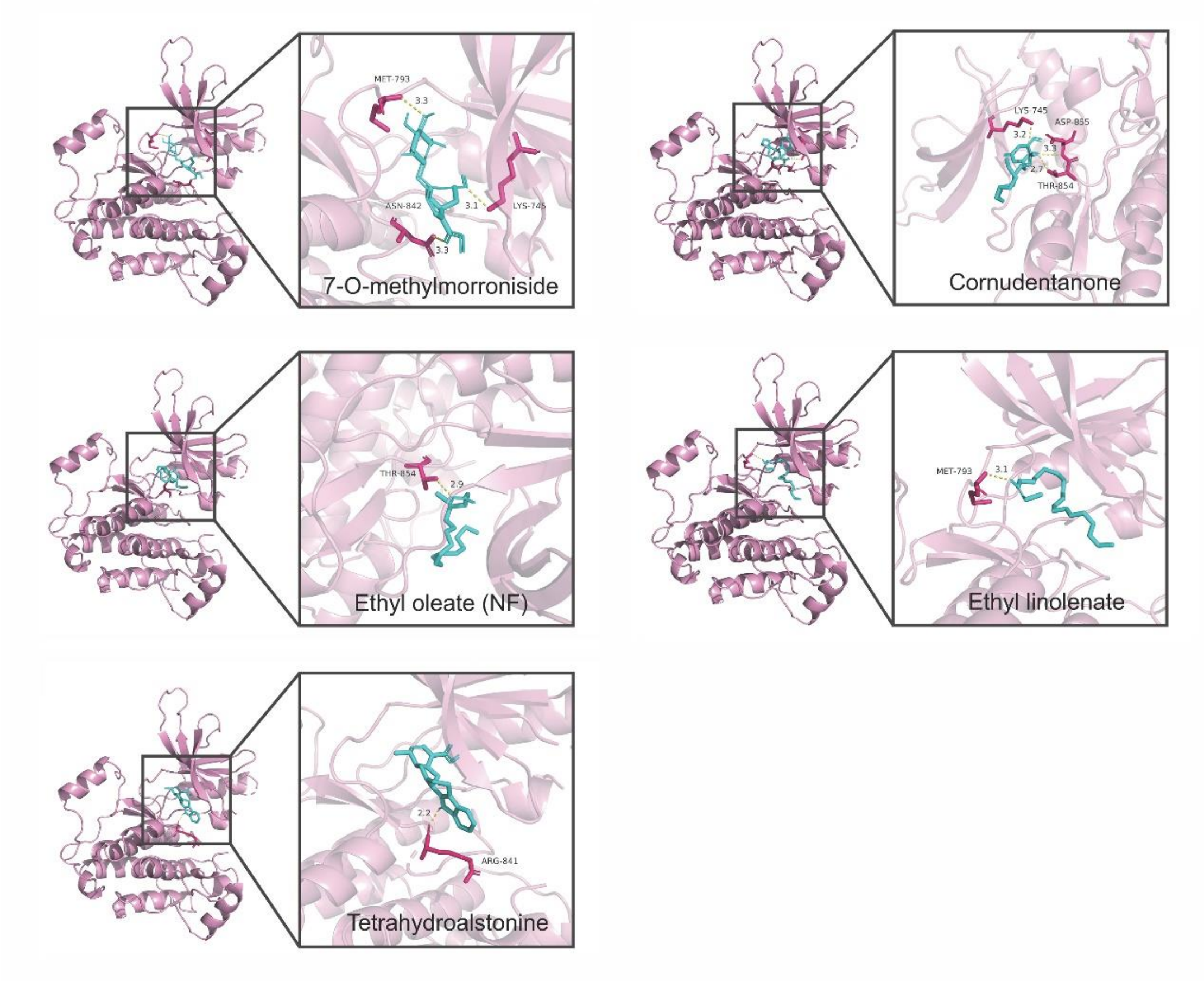

**Fig. S3. Docking schematic diagram drawn by Pymol to show the potential direct binding of ingredients of *Cornus officinalis* and EGFR.**

## Supplementary Table

**Table S1. The sequences of the primers used for qRT-PCR.**

| Name | Forward sequence | Reverse sequence |
|---|---|---|
| U6-F | CTCGCTTCGGCAGCACA | CTCGCTTCGGCAGCACA |
| U6-R | AACGCTTCACGAATTTGCGT | AACGCTTCACGAATTTGCGT |
| h- miR-103a-RT | GTCGTATCCAGTGCAGGGTCCGAGG TATTCGCACTGGATACGACTCATAG | GTCGTATCCAGTGCAGGGTCCGAGG TATTCGCACTGGATACGACTCATAG |
| h- miR-103a-F | GCGAGCAGCATTGTACAGGG | GCGAGCAGCATTGTACAGGG |
| h- miR-103a-R | AGTGCAGGGTCCGAGGTAT | AGTGCAGGGTCCGAGGTAT |
| h-ACTIN-F | CAAGCAGAAAACATGCCCGT | CAAGCAGAAAACATGCCCGT |
| h-ACTIN-R | CCGATCCACACCGAGTATTTG | CCGATCCACACCGAGTATTTG |
| h-IL-1β-F | CAAGCAGAAAACATGCCCGT | CAAGCAGAAAACATGCCCGT |
| h-IL-1β-R | AGCACAGGACTCTCTGGGTA | AGCACAGGACTCTCTGGGTA |
| h-TNF-α-F | GAACTCACTGGGGCCTACA | GAACTCACTGGGGCCTACA |
| h-TNF-α-R | GCTCCGTGTCTCAAGGAAGT | GCTCCGTGTCTCAAGGAAGT |
| h-Occludin-F | GCAAAGTGAATGACAAGCGG | GCAAAGTGAATGACAAGCGG |
| h-Occludin-R | CACAGGCGAAGTTAATGGAAG | CACAGGCGAAGTTAATGGAAG |
| h-ZO-1-F | CGAAGGAGTTGAGCAGGAAA | CGAAGGAGTTGAGCAGGAAA |
| h-ZO-1-R | ACAGGCTTCAGGAACTTGAG | ACAGGCTTCAGGAACTTGAG |
| m- miR-103a-RT | GTCGTATCGACTGCAGGGTCCGAGG TATTCGCAGTCGATACGACTCATAG | GTCGTATCGACTGCAGGGTCCGAGG TATTCGCAGTCGATACGACTCATAG |
| m- miR-103a-F | GCGAGCAGCATTGTACAGGG | GCGAGCAGCATTGTACAGGG |
| m- miR-103a-R | AGTGCAGGGTCCGAGGTATT | AGTGCAGGGTCCGAGGTATT |
| m-ACTIN-F | GGCTCCTAGCACCATGAAGA | GGCTCCTAGCACCATGAAGA |
| m-ACTIN-R | ACTCCTGCTTGCTGATCCAC | ACTCCTGCTTGCTGATCCAC |
| m-IL-1β-F | CCTGAACTCAACTGTGAAATGCC | CCTGAACTCAACTGTGAAATGCC |
| m-IL-1β-R | CAGCTTCTCCACAGCCACAATGAG | CAGCTTCTCCACAGCCACAATGAG |
| m-TNF-α-F | TACTGAACTTCGGGGTGATTGGTCC | TACTGAACTTCGGGGTGATTGGTCC |
| m-TNF-α-R | CAGCCTTGTCCCTTGAAGAGAACC | CAGCCTTGTCCCTTGAAGAGAACC |
| m-Occludin-F | TGAAAGTCCACCTCCTTACAGA | TGAAAGTCCACCTCCTTACAGA |
| m-Occludin-R | CCGGATAAAAAGAGTACGCTGG | CCGGATAAAAAGAGTACGCTGG |
| m-ZO-1-F | GCCAGAGAAAAGTTGGCAAG | GCCAGAGAAAAGTTGGCAAG |
| m-ZO-1-R | TTGGATACCACTGCGCATAA. | TTGGATACCACTGCGCATAA. |